\documentclass[11pt,letterpaper]{article}

\pdfoutput=1 
\usepackage{jheppub}
\usepackage{amsmath}
\usepackage{amsfonts}
\usepackage{amssymb}
\usepackage{graphicx}
\usepackage{epsfig}
\usepackage{natbib}

\def\ba{\begin{eqnarray}}
\def\ea{\end{eqnarray}}
\def\bea{\begin{eqnarray}}
\def\eea{\end{eqnarray}}
\def\be{\begin{equation}}
\def\ee{\end{equation}}

\def\({\left(}
\def\){\right)}
\def\[{\left[}
\def\]{\right]}

\title{Cosmological Consequences of Initial State Entanglement}

\author[a]{Andreas Albrecht,}
\author[a]{Nadia Bolis,}
\affiliation[a]{University of California at Davis, Department of Physics, One Shields Ave, Davis CA 95616 USA}
\author[b]{R.~Holman}
\affiliation[b]{Physics Department, Carnegie Mellon University, Pittsburgh PA 15213 USA}
\emailAdd{ajalbrecht@ucdavis.edu}
\emailAdd{nbolis@ucdavis.edu}
\emailAdd{rh4a@andrew.cmu.edu}

\abstract{We explore the cosmological consequences of having the
  fluctuations of the inflaton field entangled with those of another
  scalar, within the context of a toy model consisting of
  non-interacting, minimally coupled scalars in a fixed de Sitter
  background. We find that despite the lack of interactions in the
  Lagrangian, the initial state entanglement modifies the mode
  equation for the inflaton fluctuations and thus can induce changes
  in cosmological observables. These effects are examined for a
  variety of choices of masses and we find that they can be consistent
  with the requirement that the back reaction of the modified state
  not affect the inflationary phase while still giving rise to
  observable effects in the power spectrum. Our results suggest that
more realistic extensions of the ideas explored here beyond the simple
  toy model may lead to interesting observable effects.} 

\begin{document}

\maketitle

\section{Introduction\label{sec:intro}}
Modern cosmological data (for example~\cite{Ade:2013uln}) seem to
indicate that the standard inflationary paradigm consisting of a
single scalar inflaton undergoing slow-roll dynamics gives an
excellent description of the very early Universe. In particular, the
metric perturbations that give rise to CMB temperature fluctuations
and drive the growth of large scale structure are thought to have
arisen from quantum fluctuations in this field with the exponential
growth of the scale factor stretching the wavelengths of these
fluctuations from the micro to the macro scale. 

It's worth assessing our understanding of this process, and in
particular, how the initial quantum state of the fluctuations is
chosen. The scalar metric perturbations are encoded in the gauge
invariant variable $\zeta(\vec{x}, t)$ which can be thought of as the
local change in the number of e-folds of inflation. This field is then
quantized by choosing a particular solution to the relevant mode
equations and using this choice to define the Fock space vacuum
state. The standard lore then has it that since at short distances
space-time looks flat, we should choose the linear combinations of
solutions to the mode equation that match the flat space vacuum mode
functions in the short distance limit. This is how the Bunch Davies
(BD) state\cite{Bunch:1978yq} is defined. However, it is certainly
conceivable that as we go to shorter distances, other dynamical
effects could make themselves felt; for example, the inflaton could be
a composite, much like the pion, and that below a certain distance
scale $\Lambda^{-1}$ inflaton dynamics would be that of its
constituents. Arguing that the quantum state of the composite would
still be that of an undisturbed scalar would require that {\em all}
the dynamics of the formation of the composite field would be
adiabatic to an improbable extent.  More generally, until we have a
full theory of the inflaton, including its UV completion, it would be
difficult to make any certain assumptions about its vacuum
state\cite{Martin:2000xs,Danielsson:2002kx,Kaloper:2002uj,Collins:2005nu,Collins:2006bg}. An
interesting and widely discussed specific case is when inflation for
our observed ``pocket universe'' starts with a tunneling event (as
discussed for example in~\cite{Albrecht:2011yg}) which could
lead to interesting observable phenomena if the inflation within the
pocket universe is sufficiently short. Furthermore, tensions do
exist in the current data (see for example Appendix B
of~\cite{Ade:2013zuv}), which ultimately could require some
modification to the simplest picture for a resolution. 

Given these open questions, a more fruitful approach to the initial quantum state of
the inflaton might be an effective one. We should consider a more
generally parametrized state and use the new parameters as measures of
our ignorance of the process that sets the initial state. Then we can
use the available data to constrain the deviations this state might
exhibit from the BD state. It is with this philosophy in mind that we
discuss the following interesting possibility for the inflaton initial
quantum state.  

In most extensions of the standard model of particle physics, scalar
degrees of freedom other than the Higgs make an appearance. In
particular, in string theory compactifications many other scalar
fields can appear as moduli of the compactification. Let us assume
then that there is at least one other scalar field in addition to the
inflaton. Since the interactions between these fields need to be
suppressed at least to the extent that the second scalar does not ruin
the slow-roll properties of the inflaton, it would not be unreasonable
to choose the quantum state describing these fields to be the tensor
product of the inflaton state and that of the other field. However,
there is no reason we could not consider a more general initial state
such as an {\em entangled} one.  

Our goal in this work is to consider such a state and ask how the
entanglement might affect cosmological observables, such as the power
spectrum and other correlation functions of the inflaton. To simplify
matters, we will consider two  scalar fields, $\Phi$, which we will
take to be the ``inflaton'', and the non-inflaton field $\chi$
propagating in a {\em fixed} de Sitter background. A more realistic
calculation, which we defer to later work, would be to use the gauge
invariant curvature fluctuation field $\zeta$ and entangle it with the
other spectator field as well as allowing for a quasi-de Sitter
space-time. For now, our toy model will at least allow us to
understand what changes to expect in terms of inflationary
observables. 

The natural framework to use to allow for the input of entanglement
between the fields is the Schr\"odinger functional approach where we
solve the functional  Schr\"odinger equation for the wave-functional
$\Psi$ describing the evolution of the quantum state which then allows
us to compute all correlation functions using $\Psi$. Whereas this
would be quite a difficult undertaking in general, we can reduce the
degree of difficulty by taking the action for the two fields to be
completely free: there are no self-interactions, nor will we allow any
cross couplings between the fields. The only way the two fields know
about each other is through the entanglement in their joint initial
state. We can then decouple the different wave numbers for each field
mode and solve the ensuing quantum mechanical Schr\"odinger equations
for each mode wave-function $\psi_{\vec{k}}$, where $\vec{k}$ is the
comoving wave number of each mode. The effects of interactions could
then be put in perturbatively, as usual (a step we save for a future
paper).  

In the next section, we will describe the functional Schr\"odinger formalism as it applies to our problem. We then compute the wave-functional for the entangled system and understand how to trace out the non-inflaton degree of freedom to arrive at the inflaton density matrix. Given the inflaton density matrix we calculate the corrections to the power spectrum for some simple choices of entanglement parameters. We then conclude with a discussion of further directions one could take this work in. 

\section{\label{sec:schrod} Schr\"odinger Picture Field Theory: The Set-up}

There have been a number of works on the use of Schr\"odinger picture field theory in inflationary settings\cite{Boyanovsky:1993xf,Anderson:2005hi,Freese:1984dv}, so we just describe the salient points in this section. 

As discussed in the introduction, we will consider two fields $\Phi, \chi$ coupled to gravity and with no other interactions. Their action is 
\begin{eqnarray}
\label{eq:action}
& & S= \frac{1}{2}\int d^4 x\ a^4(\eta)\left[\frac{1}{a^2(\eta)}\left(\Phi^{\prime}(\eta, \vec{x})^2-\left(\nabla \Phi(\eta, \vec{x})\right)^2\right) -m_{\Phi}^2\ \Phi(\eta, \vec{x})^2\right .+\nonumber\\
& & \left . \frac{1}{a^2(\eta)}\left(\chi^{\prime}(\eta, \vec{x})^2-\left(\nabla \chi(\eta, \vec{x})\right)^2\right) -m_{\chi}^2\ \chi(\eta, \vec{x})^2\right],
\end{eqnarray}
where $a(\eta)$ is the (conformal time) scale factor of the background FRW space-time. Since the background de Sitter space has constant curvature, we allow for the possibility of non-minimal coupling to the curvature scalar by changes in the mass terms of each field.

We will need the Hamiltonian in order to be able to set up a functional Schr\"odinger equation for this system. This is easy to obtain and is given by
\begin{eqnarray}
\label{eq:ham}
& & H= \int d^3 x\ \left[\frac{\pi_{\Phi}^2}{2 a^2(\eta)}+\frac{1}{2} a^2(\eta) \left(\left(\nabla \Phi(\eta, \vec{x})\right)^2+ a^2(\eta) m_{\Phi}^2\ \Phi(\eta, \vec{x})^2\right)+\right .\nonumber\\
& & \left . \frac{\pi_{\chi}^2}{2 a^2(\eta)}+\frac{1}{2} a^2(\eta) \left(\left(\nabla \chi(\eta, \vec{x})\right)^2+ a^2(\eta) m_{\Phi}^2\ \chi(\eta, \vec{x})^2\right)\right],
\end{eqnarray}
where $\pi_{\Phi},\ \pi_{\chi}$ are the canonically conjugate momenta to $\Phi,\ \chi$ respectively. 

It will be more useful to have the Hamiltonian written in terms of the comoving spatial momentum modes, where we are taking the FRW space-time to have flat spatial sections. We decompose $\Phi$ as
\begin{equation}
\label{eq:modedecomp}
\Phi(\eta, \vec{x}) = \sum_{\vec{k}} \frac{\phi_{\vec{k}}}{\sqrt{V}} e^{-i \vec{k}\cdot \vec{x}},\quad \pi_{\Phi}(\eta, \vec{x}) = \sum_{\vec{k}} \frac{\pi_{\Phi, {\vec{k}}}}{\sqrt{V}} e^{-i \vec{k}\cdot \vec{x}},
\end{equation}
with a similar decomposition for $\chi$. Note that we are using box normalized modes with the $V$ being the comoving volume of the spatial box. In terms of these modes we have
\begin{eqnarray}
\label{eq:modehams}
& &  H = H_{\Phi} + H_{\chi}\nonumber\\
& & H_{\Phi} = \sum_{\vec{k}} H_{\Phi, \vec{k}},\quad H_{\Phi, \vec{k}} = \frac{\pi_{\Phi, \vec{k}} \pi_{\Phi, -\vec{k}}}{2 a^2(\eta)} + \frac{1}{2} a^2(\eta)\left(k^2 + m_{\Phi}^2 a^2(\eta)\right) \phi_{\vec{k}} \phi_{-\vec{k}}\nonumber\\
& &  H_{\chi} = \sum_{\vec{k}} H_{\chi, \vec{k}},\quad H_{\chi, \vec{k}} = \frac{\pi_{\chi, \vec{k}} \pi_{\chi, -\vec{k}}}{2 a^2(\eta)} + \frac{1}{2} a^2(\eta)\left(k^2 + m_{\chi}^2 a^2(\eta)\right) \chi_{\vec{k}} \chi_{-\vec{k}}
\end{eqnarray}

In the Schr\"odinger picture, the state is represented by the wave function (more generally a density matrix) $\Psi = \Psi\left[\left\{\phi_{\vec{k}}\right\}, \left\{\chi_{\vec{k}}\right\}; \eta\right]$ and the momentum operators act on this wave function as dictated by the canonical commutation relations:
$$
\pi_{\Phi, \vec{k}} \rightarrow -i\frac{\delta}{\delta \phi_{-\vec{k}}},\quad \pi_{\chi, \vec{k}} \rightarrow -i\frac{\delta}{\delta \chi_{-\vec{k}}}.
$$

The absence of interactions in our Hamiltonian allows us to factorize the wave function in terms of wave functions for each mode
\begin{equation}
\label{eq:psifactor}
\Psi\left[\left\{\phi_{\vec{k}}\right\}, \left\{\chi_{\vec{k}}\right\}; \eta\right] = \prod_{\vec{k}} \psi_{\vec{k}} \left[\phi_{\vec{k}}, \chi_{\vec{k}}; \eta\right]
\end{equation}
As mentioned in the introduction, at this point one might also want to factor $\psi_{\vec{k}}$ into pieces only depending on $\phi_{\vec{k}},\ \chi_{\vec{k}}$ separately, since there are no cross interactions. We will forgo this in order to allow for entanglement in the joint quantum state. 

A quadratic Hamiltonian begs for a Gaussian ansatz so we will write
\begin{equation}
\label{eq:Gaussansatz}
\psi_{\vec{k}} \left[\phi_{\vec{k}}, \chi_{\vec{k}}; \eta\right] = N_k(\eta) \exp\left[-\frac{1}{2}\left(A_k(\eta) \phi_{\vec{k}} \phi_{-\vec{k}}+B_k(\eta) \chi_{\vec{k}} \chi_{-\vec{k}}+C_k(\eta)\left(\phi_{\vec{k}} \chi_{-\vec{k}}+\chi_{\vec{k}} \phi_{-\vec{k}}\right)\right)\right].
\end{equation}
Here $k$ is the magnitude of the wave vector $\vec{k}$ and the entanglement is put in by demanding that $C_k(\eta_0)\neq 0$. This form of the wave function is automatically invariant under spatial translations and rotations. 

The functional Schr\"odinger equation factorizes into an infinite number of ordinary Schr\"odinger equations, one for each mode:
\begin{equation}
\label{eq:seqnmode}
i\partial_{\eta} \psi_{\vec{k}} \left[\phi_{\vec{k}}, \chi_{\vec{k}}; \eta\right]=\left(H_{\Phi, \vec{k}}+H_{\chi, \vec{k}}\right)\psi_{\vec{k}} \left[\phi_{\vec{k}}, \chi_{\vec{k}}; \eta\right].
\end{equation}
Inserting our ansatz eq.(\ref{eq:Gaussansatz}) into the above equation and then matching the powers of the field modes gives us the following equations for the normalization $N_k(\eta)$ and the kernels $A_k(\eta),\ B_k(\eta),\ C_k(\eta)$ (where the primes denote conformal time derivatives):
\begin{eqnarray}
\label{eq:kernels}
& & i\frac{N_k^{\prime}}{N_k} = \frac{\left(A_k+B_k\right)}{2 a^2(\eta)} \nonumber\\
& & i A_k^{\prime} = \frac{A_k^2+C_k^2}{a^2(\eta)} -\Omega_{\Phi, k}^2 a^2(\eta),\quad \Omega_{\Phi, k}^2\equiv k^2 + m_{\Phi}^2 a^2(\eta)\nonumber\\
& &  i B_k^{\prime} = \frac{B_k^2+C_k^2}{a^2(\eta)} -\Omega_{\chi, k}^2 a^2(\eta),\quad \Omega_{\chi, k}^2\equiv k^2 + m_{\chi}^2 a^2(\eta)\nonumber\\
& &  i\frac{C_k^{\prime}}{C_k}=  \frac{\left(A_k+B_k\right)}{a^2(\eta)}.
\end{eqnarray}
We can see from the last equation that if $C_k(\eta_0)$ vanished, then $C_k(\eta)$ would vanish identically and our state would then factorize. We also note that the equations for $A_k,\ B_k$ are of the Ricatti form and can be converted into {\em linear}, second order equations by writing 
\begin{equation}
i A_k(\eta) = a^2(\eta)\left(\frac{f_k^{\prime}(\eta)}{f_k(\eta)}- \frac{a^{\prime}(\eta)}{a(\eta)}\right),\quad i B_k(\eta) = a^2(\eta)\left(\frac{g_k^{\prime}(\eta)}{g_k(\eta)}- \frac{a^{\prime}(\eta)}{a(\eta)}\right).
\end{equation}
The resulting equations for $f_k(\eta),\ g_k(\eta)$ are
\begin{eqnarray}
\label{eq:modefuncts}
& & f_k^{\prime \prime} + \left(\Omega_{\Phi, k}^2-\frac{a^{\prime \prime}(\eta)}{a(\eta)}\right) f_k = \frac{C_k(\eta)^2}{a^4(\eta)}f_k\nonumber\\
& & g_k^{\prime \prime} + \left(\Omega_{\chi, k}^2-\frac{a^{\prime \prime}(\eta)}{a(\eta)}\right) g_k = \frac{C_k(\eta)^2}{a^4(\eta)}g_k
\end{eqnarray}
Furthermore, the equation for $C_k$ yields the relation
\begin{equation}
\label{eq:solveforc}
\frac{C_k (\eta)}{a^2(\eta)} = \frac{\lambda_k}{f_k(\eta) g_k(\eta)},
\end{equation}
where $\lambda_k$ is a constant. 

We have traded a set of first order, non-linear equations for second order linear ones. The question then arises as to where the extra initial conditions required to fully solve the latter set come from. The clue to solving this puzzle comes from the fact that the kernels $A_k(\eta),\ B_k(\eta)$ only depend on the {\em ratios} $f_k^{\prime}\slash f_k,\ g_k^{\prime}\slash g_k$. What this means operationally is that of the two integration constants required to specify a solution of eq.(\ref{eq:modefuncts}) only their ratio is physical. We can use this freedom to specify the Wronskians $W[f_k, f_k^*],\ 
W[g_k, g_k^*]$, which are time independent. We will take both of these Wronskians to equal $-i$ so that we can write
\begin{equation}
\label{eq:initconds1}
f_k^{\prime}(\eta_0) = \left(\frac{i A_k(\eta_0)}{a^2(\eta_0)}+\left . \frac{a^{\prime}}{a}\right |_{\eta=\eta_0}\right) f_k(\eta_0),\ \quad \left |f_k(\eta_0)\right |^2 = \frac{a^2(\eta_0)}{2 A_{k R}(\eta_0)},
\end{equation}
where $A_{k R}$ is the real part of $A_k$ with a similar equation for $g_k$. The second relation follows from the Wronskian condition while the first just comes from Ricatti equation change of variables evaluated at the initial time. We see then that to fully specify a solution for our mode functions, we need to specify $A_k(\eta_0)$.

Before we turn to this task, it's worth noting that with our choice of Wronskian normalization, the mode functions $f_k,\ g_k$ have mass dimension $-1\slash 2$ as do the modes $\phi_{\vec{k}},\ \chi_{\vec{k}}$. This implies that the kernels have mass dimension $+1$ and thus that the parameter $\lambda_k$ is dimensionless. 

There is one more constraint that must be enforced for the wave function arrived at above to describe a physically allowed quantum state: it must be normalizable, i.e. 
\begin{equation}
\label{eq:normalization1}
\int {\cal D}^2\phi_{\vec{k}}\ {\cal D}^2\chi_{\vec{k}} \left |\psi_{\vec{k}}\right |^2  \left |\psi_{-\vec{k}}\right |^2<\infty.
\end{equation}
The functional measure ${\cal D}^2\phi_{\vec{k}}\ {\cal D}^2\chi_{\vec{k}}$ includes the contribution arising from $\vec{k}\rightarrow -\vec{k}$ since the corresponding modes are complex conjugates of one another. That is to say, ${\cal D}^2\phi_{\vec{k}} \equiv {\cal D}{\rm Re}\phi_{\vec{k}}\ {\cal D}{\rm Im}\phi_{\vec{k}}$ and likewise for ${\cal D}^2\chi_{\vec{k}}$.  This fact also accounts for the extra term $\left |\psi_{-\vec{k}}\right |^2$ in the integrand. Putting everything together, eq.(\ref{eq:normalization1}) becomes
\begin{equation}
\label{eq:normalization2}
\int {\cal D}^2\phi_{\vec{k}}\ {\cal D}^2\chi_{\vec{k}} \exp\left[-2\left(\begin{array}{cc}\phi_{\vec{k}} & \chi_{\vec{k}}\end{array}\right)\left(\begin{array}{cc}A_{k R} & C_{k R} \\C_{k R} & B_{k R}\end{array}\right) \left(\begin{array}{c}\phi_{-\vec{k}} \\\chi_{-\vec{k}}\end{array}\right)\right]<\infty
\end{equation}
This integral can be done in the usual way and it is proportional to $1\slash \det {\cal O}_k$ where $ {\cal O}_k$ is the matrix appearing in the exponential in eq.(\ref{eq:normalization2}). In order for the integral to be finite, both of the eigenvalues of  ${\cal O}_k$ must be positive which implies the requirement: $A_{k R} B_{k R}-C_{k R}^2>0$. We can rewrite this in terms of the mode functions as
\begin{equation}
\label{eq:normconstraint}
A_{k R} B_{k R}-C_{k R}^2 = \frac{a^4(\eta)}{4 \left | f_k(\eta)\right|^2 \left | g_k(\eta)\right|^2}\left(1-4 \left| \lambda_k\right |^2 \cos^2 \left(\rho_k -\theta_{f k}(\eta)-\theta_{g k}(\eta)\right)\right)>0,
\end{equation}
where we have defined $\rho_k,\ \theta_{k f},\ \theta_{k g}$ via 
$$
\lambda_k = \left| \lambda_k\right | e^{i \rho_k},\quad f_k(\eta) = \left | f_k(\eta)\right|  e^{i \theta_{k f}},\quad  g_k(\eta) = \left | g_k(\eta)\right|  e^{i \theta_{k g}}.
$$
We see that the normalization constraint can be satisfied as long as $\left| \lambda_k\right |<1\slash 2$. 

What should we do about the initial values of the kernels $A_k,\ B_k$? We would like to compare the effects of having an entangled state to the standard picture of inflation, i.e. where the initial inflaton state is chosen to be the BD state. This would correspond to setting $A_k(\eta_0)$ equal to $A^{\rm BD}_k(\eta_0)$ where 
\begin{equation}
\label{eq:BDkernel}
i A^{\rm BD}_k(\eta)= a^2(\eta)\left(\frac{f_k^{\rm BD \prime}(\eta)}{f^{\rm BD}_k(\eta)}- \frac{a^{\prime}(\eta)}{a(\eta)}\right),
\end{equation}
with 
\begin{equation}
\label{eq:BDmode}
f^{\rm BD}_k(\eta) = \frac{\sqrt{\pi}}{2} \sqrt{-\eta} H_{\nu_{\Phi}}^{(1)}(-k \eta),\quad \nu_{\Phi} = \sqrt{\frac{9}{4}-\frac{m_{\Phi}^2}{H_I^2}}
\end{equation}
where $H_{\nu_{\Phi}}^{(1)}(-k \eta)$ is the Hankel function of the first kind. We will also take $\chi$ to be in its BD vacuum which implies that a similar story to the one above applies to $B_k$. 

Putting all this together, we need to solve the following equations:
\begin{eqnarray}
\label{eq:finaleqs}
& & f_k^{\prime \prime} + \left(\Omega_{\Phi, k}^2-\frac{a^{\prime \prime}(\eta)}{a(\eta)}\right) f_k = \frac{\lambda_k^2}{f_k g_k^2}\nonumber\\
& & g_k^{\prime \prime} + \left(\Omega_{\chi, k}^2-\frac{a^{\prime \prime}(\eta)}{a(\eta)}\right) g_k = \frac{\lambda_k^2}{f_k^2 g_k},
\end{eqnarray}
subject to the initial conditions
\begin{equation}
\label{eq:finalinitconds}
f_k(\eta_0)= f^{\rm BD}_k(\eta_0),\quad f^{\prime}_k(\eta_0) = f^{\rm BD \prime}_k(\eta_0);\quad g_k(\eta_0)= g^{\rm BD}_k(\eta_0),\quad g^{\prime}_k(\eta_0) = g^{\rm BD \prime}_k(\eta_0).
\end{equation}

\subsection{\label{subsec:pert} Perturbative Solution}

We will consider numerical solutions of eqs.(\ref{eq:finaleqs}) below, but some aspects of the solutions can be uncovered by a perturbative approach. We already know that BD modes give a good account of the data, so that any deviations, which are parametrized by the dimensionless coupling $\lambda_k$ must be small. It would make sense then to use $\lambda_k$ as a perturbative parameter in order to solve eqs.(\ref{eq:finaleqs},\ref{eq:finalinitconds}). For small values of $\lambda_k$ ($|\lambda_k|\ll1\slash 2$) we should expect this perturbative solution to match the actual one quite well. However, there might be some interesting effects that appear when $\lambda_k$ is near its largest allowed value at which point a numerical solution is required.

Let's write
\begin{equation}
f_k(\eta) = f_k^{(0)}(\eta)\left(1+ \lambda_k^2 {\cal F}_k(\eta)\right),\quad g_k(\eta) = g_k^{(0)}(\eta)\left(1+ \lambda_k^2 {\cal G}_k(\eta)\right),
\end{equation}
and then insert this in eq.(\ref{eq:finaleqs}), matching powers of $\lambda_k$. We find that $f_k^{(0)}(\eta),\ g_k^{(0)}(\eta)$ satisfy the same equation satisfied by the BD modes and the initial conditions in eq.(\ref{eq:finalinitconds}) then force $f_k^{(0)}(\eta)=f_k^{BD}(\eta),\ g_k^{(0)}(\eta)=g_k^{BD}(\eta)$. The ${\cal O}(\lambda_k^2)$ terms then give:
\begin{eqnarray}
\label{eq:perteqs}
& & {\cal F}_k^{\prime \prime}(\eta)+ 2 \frac{f_k^{(0) \prime}(\eta)}{f_k^{(0)}(\eta)}  {\cal F}_k^{\prime} = \frac{1}{f_k^{(0) 2}(\eta) g_k^{(0) 2}(\eta)}\nonumber\\
& & {\cal G}_k^{\prime \prime}(\eta)+ 2 \frac{g_k^{(0) \prime}(\eta)}{g_k^{(0)}(\eta)}  {\cal G}_k^{\prime} = \frac{1}{f_k^{(0) 2}(\eta) g_k^{(0) 2}(\eta)},
\end{eqnarray}
together with the initial conditions
\begin{equation}
\label{eq:pertinitconds}
{\cal F}_k(\eta_0)={\cal G}_k(\eta_0)=0,\quad {\cal F}_k^{\prime}(\eta_0)={\cal G}_k^{\prime}(\eta_0)=0.
\end{equation}
Eqs.(\ref{eq:perteqs}) admit an integrating factor and using the initial conditions we then find:
\begin{eqnarray}
\label{eq:finalperts}
& & {\cal F}_k(\eta)= \int_{\eta_0}^{\eta} d\eta_1\ \frac{1}{f_k^{BD}(\eta_1)^2}  \int_{\eta_0}^{\eta_1} d\eta_2\ \frac{1}{g_k^{BD}(\eta_2)^2},  \nonumber\\
& & {\cal G}_k(\eta)= \int_{\eta_0}^{\eta} d\eta_1\ \frac{1}{g_k^{BD}(\eta_1)^2}  \int_{\eta_0}^{\eta_1} d\eta_2\ \frac{1}{f_k^{BD}(\eta_2)^2}.
\end{eqnarray}

Setting $f_k^{BD}(\eta) = g_k^{BD}(\eta)$  allows for a great simplification in the expressions for ${\cal F}_k, {\cal G}_k$. If we interchange the order of integration in eq.(\ref{eq:finalperts}), relabel $\eta_1\leftrightarrow \eta_2$ and add the results, we find 
\begin{equation}
\label{eq:equalmodeperts}
{\cal F}_k(\eta) = {\cal G}_k(\eta)= \frac{1}{2} \left(\int_{\eta_0}^{\eta} \frac{d\eta^{\prime}}{ f_k^{BD}(\eta^{\prime})^{2}}\right)^2.
\end{equation}

\section{\label{sec:infcosmoobs} Inflaton Cosmological Observables}

We now turn to the task of computing what would be the important inflationary cosmological observables if $\Phi$ did indeed measure primordial curvature perturbations. While we could do this directly from the wave function of eq.(\ref{eq:Gaussansatz}), we will find it instructive to first construct the reduced density matrix for $\Phi$ and use it to compute the power spectrum. Having the inflaton density matrix at hand will also allow us to impose the constraints coming due to backreaction of the energy density associated with this state.

\subsection{\label{sunsec:infstate}  The Inflaton Density Matrix}

The reduced density matrix, $\rho_{\Phi}$, for $\Phi$ is found by tracing out the $\chi$ degrees of freedom from the wave function in eq.(\ref{eq:Gaussansatz}). It has the following matrix elements in the Schr\"odinger picture:
\begin{equation}
\label{eq:reddensmat}
\langle \phi\left |\rho_{\Phi}(\eta)\right | \tilde{\phi}\rangle \equiv \int {\cal D}\chi\ \langle \phi, \chi \left |\Psi(\eta)\right . \rangle \langle \Psi(\eta) \left | \tilde{\phi}, \chi \right . \rangle.
\end{equation}
Due to the fact that the original wave function was the product of the wave functions for each wave number $\vec{k}$, the above expression also factorizes into the product of density matrix elements:
\begin{equation}
\label{eq:factdensmat}
\rho_{\vec{k}}\left[\phi_{\vec{k}},\tilde{\phi}_{\vec{k}};\eta\right] = \int {\cal D}^2\chi_{\vec{k}}\ \langle \phi_{\vec{k}}, \chi_{\vec{k}} \left |\Psi_{\vec{k}}(\eta)\right . \rangle \langle \Psi_{\vec{k}}(\eta) \left | \tilde{\phi}_{\vec{k}}, \chi_{\vec{k}} \right . \rangle.
\end{equation}
The integrals can all be done easily and in the end we find
\begin{equation}
\label{eq:infdensmat}
\rho_{\vec{k}}\left[\phi_{\vec{k}},\tilde{\phi}_{\vec{k}};\eta\right]=\frac{2}{\pi} \left(A_{k R}-\frac{C_{k R}^2}{B_{k R}}\right) \exp\left[-\gamma_k \phi_{\vec{k}}\phi_{-\vec{k}}-\gamma_k^* \tilde{\phi}_{\vec{k}} \tilde{\phi}_{-\vec{k}}+\beta_k\left(\phi_{\vec{k}}\tilde{\phi}_{-\vec{k}}+\phi_{-\vec{k}}\tilde{\phi}_{\vec{k}}\right)\right]
\end{equation}
where 
\begin{equation}
\label{eq:densmatpars}
\gamma_k \equiv A_k-\frac{C_k^2}{2 B_{k R}},\quad \beta_k \equiv \frac{\left | C_k\right |^2}{2 B_{k R}}
\end{equation}
and the prefactor can be found from demanding the trace of the density matrix be unity. 

As expected, we see that the reduced density matrix corresponds to a mixed state due to the loss of information about $\chi$ and that the mixing is controlled by the entanglement parameter $C_k$. 

\subsection{\label{subset:constraints} The Backreaction Constraint}

If we want inflation to actually occur, the energy density in the inflaton fluctuations must be smaller than that of the inflaton background. This translates to the constraint
\begin{equation}
\label{eq:backreaction}
\rho_{\rm flucts} \equiv \langle T^0_0 \rangle \ll M_{Pl}^2 H_I^2,
\end{equation}
where $H_I$ is the Hubble parameter during inflation and the expectation value of the stress energy is taken in our state. For a massless, minimally coupled free field the energy density $\rho_{\rm flucts}$ can be written as a sum of contributions from each mode. These latter quantities are given by
\begin{equation}
\label{eq:modeenergydens}
\rho_{\vec{k}} (\eta) = \frac{\langle \pi_{\vec{k}} \pi_{-\vec{k}}\rangle}{2 a^6(\eta)} + \frac{\Omega_{\Phi,k}^2}{2 a^2(\eta)} \langle \phi_{\vec{k}} \phi_{-\vec{k}}\rangle.
\end{equation}

The $\Phi$ two-point function can be easily computed, since the density matrix has already been written in the field representation:
\begin{eqnarray}
\label{eq:2ptfcn}
& & \langle \phi_{\vec{k}} \phi_{-\vec{k}} \rangle(\eta) \equiv {\rm Tr}\left(\rho_{\vec{k}}(\eta) \phi_{\vec{k}} \phi_{-\vec{k}} \right)=\nonumber\\
& & =\frac{2}{\pi} \left(A_{k R}-\frac{C_{k R}^2}{B_{k R}}\right) \int {\cal D}^2 \phi_{\vec{k}}\ \phi_{\vec{k}} \phi_{-\vec{k}}\ \exp\left(-2 (\gamma_{k R} - \beta_k) \phi_{\vec{k}} \phi_{-\vec{k}}\right)=\nonumber\\
& & = \frac{B_{k R}}{2\left(A_{k R} B_{k R}-C_{k R}^2\right)}.
\end{eqnarray}

We can rewrite this in terms of the modes $f_k,\ g_k$, using the definitions  in eq.(\ref{eq:normconstraint}) as 

\begin{equation}
\label{eq:final2pt}
 \langle \phi_{\vec{k}} \phi_{-\vec{k}} \rangle(\eta)= \frac{\left | f_k(\eta)\right |^2}{a^2(\eta)}\left(\frac{1}{1-4 \left|\lambda_k\right|^2 \cos^2 \left(\rho_k -\theta_{k f}-\theta_{k g}\right)}\right)
\end{equation}
The first factor, with the standard BD mode in place of $f_k(\eta)$ would be the usual result for the 2-pt function in the absence of any entanglement. The change in the mode, together with the second factor encodes all the new physics brought on by entangling the inflaton with $\chi$. 

In the Schr\"odinger picture $\langle \pi_{\vec{k}} \pi_{-\vec{k}}\rangle$ is given by 
\begin{equation}
\label{eq:momenergydens}
\langle \pi_{\vec{k}} \pi_{-\vec{k}}\rangle = \int {\cal D}^2\phi_{\vec{k}}\ \left . \left(-\frac{\delta^2}{\delta \phi_{\vec{k}} \delta \phi_{-\vec{k}}} \langle \phi_{\vec{k}} \left | \rho(\eta) \right | \tilde{\phi}_{\vec{k}}\rangle\right) \right |_{\tilde{\phi}_{\vec{k}}= \phi_{\vec{k}}}=\gamma_k -(\gamma_k-\beta_k)^2 \langle \phi_{\vec{k}} \phi_{-\vec{k}}\rangle,
\end{equation}
where $\gamma_k,\ \beta_k$ are as given in eq.(\ref{eq:densmatpars}). Using the above result that $\langle \phi_{\vec{k}} \phi_{-\vec{k}}\rangle= 1\slash (2(\gamma_{k R}-\beta_k))$ we can show that this expectation value is indeed real as befits a hermitian operator and that 
\begin{equation}
\langle \pi_{\vec{k}} \pi_{-\vec{k}}\rangle = \frac{1}{2} \left(\frac{\left| \gamma_k \right|^2-\beta_k^2}{\gamma_{k R}-\beta_k}\right). 
\end{equation}
To write this in terms of the mode functions,  set $\Theta_k(\eta)\equiv \rho_k-\theta_{k f}(\eta)-\theta_{k g}(\eta)$, to arrive at:
\begin{eqnarray}
\label{eq:parsintermsofmodes}
& & \beta_k = \left |\lambda_k\right|^2 \left(\frac{\left | f_k(\eta)\right|^2}{a^2(\eta)}\right)^{-1}\nonumber\\
& & \gamma_{k R} = \frac{1}{2} \left(\frac{\left |f_k(\eta)\right|^2}{a^2(\eta)}\right)^{-1} \left(1-2 \left |\lambda_k\right|^2 \cos 2\ \Theta_k(\eta)\right)\nonumber\\
& & \gamma_{k I} =\frac{1}{2} \left(\frac{\left |f_k(\eta)\right|^2}{a^2(\eta)}\right)^{-1}\left( \left(\frac{\left |f_k(\eta)\right|^2}{a^2(\eta)}\right)^{\prime} -2 \left |\lambda_k\right|^2 \sin 2\ \Theta_k(\eta)\right),
\end{eqnarray}
where again the prime denotes an $\eta$ derivative. Putting all these results together we find
\begin{equation}
\label{eq:mom2pt}
\langle \pi_{\vec{k}} \pi_{-\vec{k}}\rangle = \frac{1}{4} \left(\frac{\left |f_k(\eta)\right|^2}{a^2(\eta)}\right)^{-1}\left(\frac{1+\left[\partial_{\eta}\left(\frac{\left |f_k(\eta)\right|^2}{a^2(\eta)}\right)\right]^2-4 \left |\lambda_k\right|^2 \left(\cos 2\ \Theta_k(\eta)+\partial_{\eta}\left(\frac{\left |f_k(\eta)\right|^2}{a^2(\eta)}\right)\sin 2\ \Theta_k(\eta)\right)}{1-4 \left |\lambda_k\right|^2 \cos^2\ \Theta_k(\eta)}\right).
\end{equation}
We could try to bound the total energy density coming from our results, but we will be better served by using the perturbative expansion developed above. 

Taking $\rho_k=0$ as done earlier gives us the following result:
\begin{eqnarray}
\label{eq:field2ptfch}
& & \langle \phi_{\vec{k}} \phi_{-\vec{k}} \rangle(\eta)= \frac{\left | f^{BD}_k(\eta)\right |^2}{a^2(\eta)}\left(1+ \lambda_k^2\left(2 {\rm Re}\ {\cal F}_k +4 \cos^2 \left(\theta^{BD}_{k f}+\theta^{BD}_{k g}\right)\right)\right)=\nonumber\\
& &=  \langle \phi_{\vec{k}} \phi_{-\vec{k}} \rangle^{BD}(\eta)+\lambda_k^2 \frac{\left | f^{BD}_k(\eta)\right |^2}{a^2(\eta)}\left(2 {\rm Re}\ {\cal F}_k +4 \cos^2 \left(\theta^{BD}_{k f}+\theta^{BD}_{k g}\right)\right)
\end{eqnarray}
\begin{eqnarray}
\label{eq:mom2ptfcn}
& & \langle \pi_{\vec{k}} \pi_{-\vec{k}}\rangle=\langle \pi_{\vec{k}} \pi_{-\vec{k}}\rangle^{BD}+\nonumber\\ \nonumber\\
& & \frac{\lambda_k^2}{4} \left(\frac{\left |f^{BD}_k(\eta)\right|^2}{a^2(\eta)}\right)^{-1}\left(4\sin^2 (\theta^{BD}_{k f}(\eta)+\theta^{BD}_{k g}(\eta))-4 \sin 2(\theta^{BD}_{k f}(\eta)+\theta^{BD}_{k g}(\eta)) \partial_{\eta}\left(\frac{\left |f^{BD}_k(\eta)\right|^2}{a^2(\eta)}\right)\right .\nonumber\\
& & \left .+4\cos^2(\theta^{BD}_{k f}(\eta)+\theta^{BD}_{k g}(\eta)) \left(\partial_{\eta}\left(\frac{\left |f^{BD}_k(\eta)\right|^2}{a^2(\eta)}\right)\right)^2-2{\rm Re}\ {\cal F}_k(\eta)+2{\rm Re}\ {\cal F}_k(\eta) \left(\partial_{\eta}\left(\frac{\left |f^{BD}_k(\eta)\right|^2}{a^2(\eta)}\right)\right)^2\right .\nonumber\\
& & \left . +4 \left(\frac{\left |f^{BD}_k(\eta)\right|^2}{a^2(\eta)}\right)\left(\partial_{\eta}\left(\frac{\left |f^{BD}_k(\eta)\right|^2}{a^2(\eta)}\right)\right)\partial_{\eta} {\rm Re}\ {\cal F}_k(\eta)\right ),
\end{eqnarray}
where $\theta^{BD}_{k f}(\eta),\ \theta^{BD}_{k g}(\eta)$ are the relevant phases of the BD modes.

The energy density coming only from the Bunch-Davies terms will have the usual UV divergences which we will assume are dealt with by renormalizing  the BD contribution in the usual way. We will attempt to put bounds on the new contribution coming from the entanglement. In general this is a difficult task but we can get an idea of what the constraint might look like by taking both $\Phi$ and $\chi$ to be massless minimally coupled scalars, so that $\nu_{\Phi}= \nu_{\chi} = 3\slash 2$:
\begin{equation}
\label{eq:massmincoupBD}
f_k^{BD}(\eta) = g_k^{BD}(\eta) = -\frac{e^{i k \eta} (k\ \eta+i)}{\sqrt{2}  k^{3/2}\ \eta},
\end{equation}
We can extract the phases $\theta^{BD}_{k f}=\theta^{BD}_{k g}=k\eta + \tan^{-1}\left(1\slash k\eta\right)$ as well as compute the perturbative corrections in eq.(\ref{eq:equalmodeperts}) explicitly using eq.(\ref{eq:massmincoupBD}). We have
\begin{equation}
\label{eq:indefint}
\int_{\eta_0}^{\eta} \frac{d\eta^{\prime}}{ f_k^{BD}(\eta^{\prime})^{2}}= \exp(i G(k\eta))-\exp(i G(k\eta_0)),\quad G(x)\equiv -2 x + \tan^{-1} x-\tan^{-1}\frac{1}{x}.
\end{equation}
Squaring this we find 
\begin{equation}
{\cal F}_k(\eta) = \frac{1}{2}\left( \exp(2 i G(k\eta))+ \exp(2 i G(k\eta_0))-2  \exp(i (G(k\eta))+G(k\eta_0))\right), 
\end{equation}
so that 
\begin{equation}
\label{eq:calFreal}
{\rm Re}\ {\cal F}_k = \frac{1}{2}\left( \cos(2  G(k\eta))+ \cos(2  G(k\eta_0))-2  \cos(G(k\eta)+G(k\eta_0))\right)
\end{equation}

The largest contribution to the energy density comes at the earliest time $\eta_0$. Further, the contributions coming from ${\rm Re}\ {\cal F}_k$ will be oscillatory and bounded. We find that the bounds on $\lambda_k$ are not very stringent and are of the form (here $\Lambda$ is the comoving momentum cutoff):
\begin{equation}
\label{eq:backreactionconstraint}
H^4 \eta_0^4 \int_0^{\Lambda} dk\ k^3 \lambda_k^2 \ll M_{\rm Pl}^2 H^2.
\end{equation}
This integral constraint is easy enough to satisfy for a variety of reasonable choices of $\lambda_k$, such as power laws or damped exponentials.

\subsection{\label{subsec:powerspectrum} The Inflaton Power Spectrum}

A full scale investigation into the effects of initial state entanglement will have to wait for later work. In the meantime, though, we can at least examine some simple examples that might help guide us. We will first set up our mode equations in such a way as to be more amenable to numerical solutions. We will then use these solutions to compute the power spectrum 
\begin{equation}
\label{eq:dimensionlesspowerspec}
\Delta_{\Phi}^2(k) \equiv \frac{k^3}{2\pi^2} \langle \phi_{\vec{k}} \phi_{-\vec{k}} \rangle\left |_{\eta\rightarrow 0^-}\right .
\end{equation}
In standard inflation, this would have power-law behavior $\propto k^{n_s-1}$, with $n_s$ being the spectral index. 

\subsubsection{\label{subsubsec:numsetup} Numerical Set Up}

We will rewrite eqs.(\ref{eq:finaleqs},\ref{eq:finalinitconds}) in dimensionless form by defining $\tau \equiv \eta\slash \eta_0$, $q= -k\eta_0$, $\tilde{f}_q(\tau)\equiv f_k(\eta)\slash \sqrt{-\eta_0}$ and likewise for $\tilde{g}_q (\tau)$. It is easy to see that $\tau = \exp(-N(\eta))$ where $N(\eta)$ is the number of e-folds elapsed since the beginning of inflation. We can also interpret $q$ as measuring $k$ in units of the wavenumber $k_{\rm min} = -1\slash \eta_0$ which corresponds to the scale that left the de Sitter horizon at $\eta=\eta_0$. Note that the range of $\tau$ is from the initial value $\tau=1$ to zero, or more realistically, to $\tau_{\rm end}$ which would denote the end of inflation. 

In terms of these variables our equations and initial conditions become
\begin{eqnarray}
\label{eq:dimensionlesseqs}
& & \left(\frac{d^2}{d\tau^2} + \left(q^2 + \frac{\frac{1}{4}-\nu_{\Phi}^2}{\tau^2}\right)\right)\tilde{f}_q(\tau)= \frac{\lambda_q^2}{\tilde{g}_q(\tau)^2\tilde{f}_q(\tau)}\nonumber\\
& & \left(\frac{d^2}{d\tau^2} + \left(q^2 + \frac{\frac{1}{4}-\nu_{\chi}^2}{\tau^2}\right)\right)\tilde{g}_q(\tau)= \frac{\lambda_q^2}{\tilde{f}_q(\tau)^2\tilde{g}_q(\tau)}\nonumber\\
 & & \tilde{f}_q(\tau=1) = \frac{\sqrt{\pi}}{2}\ H^{(1)}_{\nu_{\Phi}}(q),\quad  \partial_{\tau} \tilde{f}_q(\tau)\left|_{\tau=1}\right . = \partial_{\tau}\left(\frac{\sqrt{\pi \tau}}{2}\ H^{(1)}_{\nu_{\Phi}}(q \tau)\right)\left|_{\tau=1}\right . \nonumber\\
 & & \tilde{g}_q(\tau=1) = \frac{\sqrt{\pi}}{2}\ H^{(1)}_{\nu_{\chi}}(q),\quad  \partial_{\tau} \tilde{g}_q(\tau)\left|_{\tau=1}\right . = \partial_{\tau} \left(\frac{\sqrt{\pi \tau}}{2}\ H^{(1)}_{\nu_{\chi}}(q \tau)\right)\left|_{\tau=1}\right .
\end{eqnarray}
 
The power spectrum becomes
\begin{equation}
\label{eq:dimlesspowerspec}
\Delta_{\Phi}^2 (q) = \left(\frac{H^2}{2\pi^2}\right) \left(\frac{q^3 \tau^2 \left|\tilde{f}_q(\tau)\right|^2}{1-4\lambda_q^2 \cos^2 \Theta_q(\tau)}\right)\equiv \left(\frac{H^2}{2\pi^2}\right)\tilde{\Delta}^2 (q),
\end{equation}
where $\Theta_q(\tau)\equiv \theta_{q \tilde{f}}(\tau)+\theta_{q \tilde{g}}(\tau)$, $\tau\rightarrow 0$ and we have taken $\lambda_q$ to be real. While the back reaction constraint of eq.(\ref{eq:backreactionconstraint}) puts bounds on the high $q$ behavior of $\lambda_q$ we will take it to be essentially constant for our numerical work. We expect that, given that to a large extent the power spectrum exhibits scale invariance at least for observationally accessible scales, and that $\lambda_q$ controls the running of the spectral index, $\lambda_q$ should not have too great a dependence on $q$ over these scales. 

\begin{figure}[htbp]
\begin{center}
\includegraphics[scale=0.6]{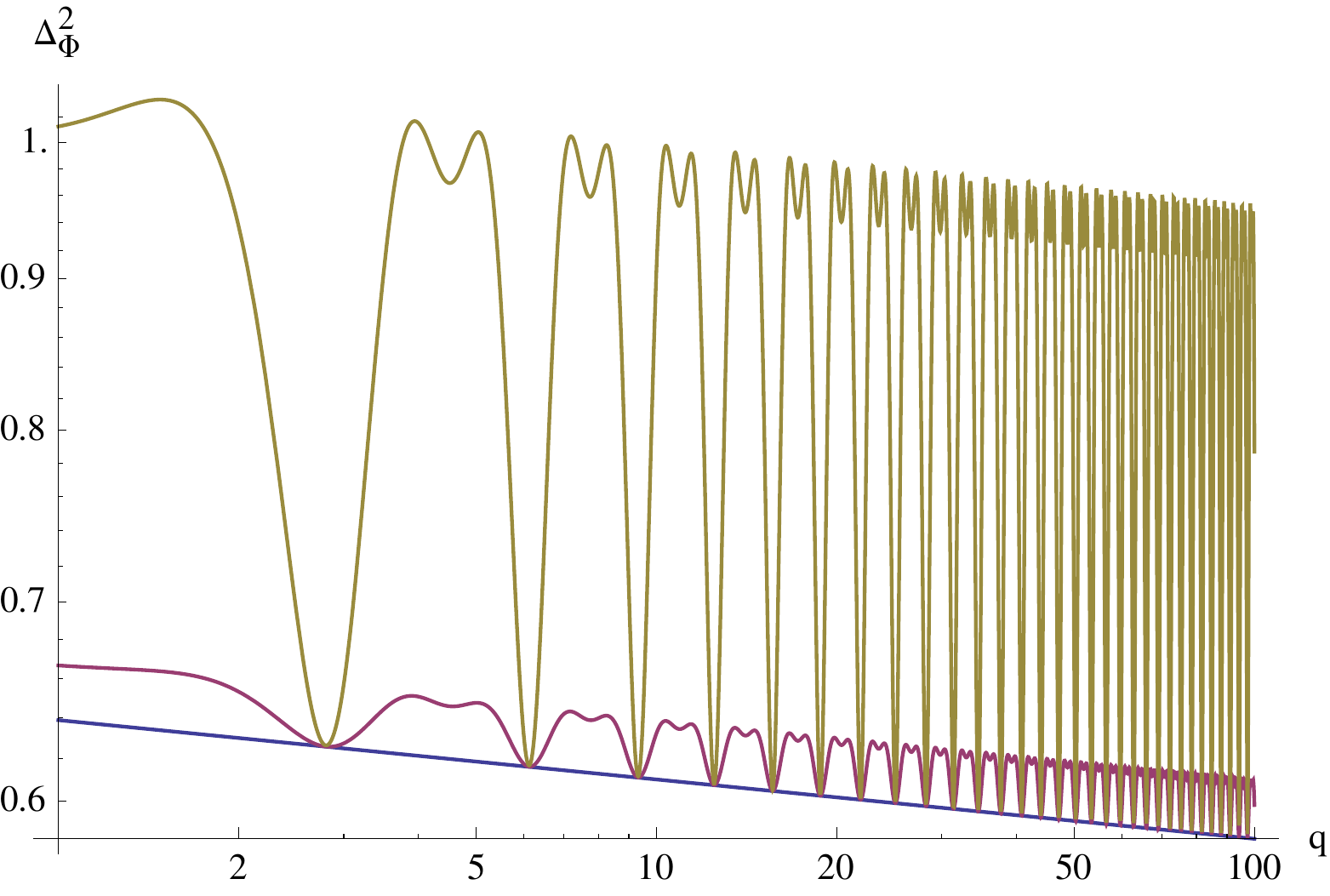}
\caption{The power $\tilde{\Delta}^2_{\Phi}(q)$ versus $q$,
  for different values of the entanglement parameter $\lambda_q = 0,
  0.1, 0.3$ using $\Phi$ nearly massless and $\chi$
  massless. The non-entangled curve is straight, and increasing the
  entanglement introduces increasing amplitudes of oscillatory behavior on
  top of the straight piece.} 
\label{fig:epsilon}
\end{center}
\end{figure}

In fig.(\ref{fig:epsilon}) we exhibit  a plot of $\tilde{\Delta}^2
(q)$ as a function of $q$ for $\lambda_q = 0, 0.1, 0.3$. We take the
inflaton field to have $\nu_{\Phi}= 3\slash 2 + \epsilon$, with
$\epsilon=0.01$ allowing for a closer relation to slow-roll inflation
in the sense of giving a red tilted spectrum, while $\nu_{\chi} =
3\slash 2$, so that $\chi$ is massless and minimally coupled. We see
that for $\lambda_q\neq 0$, the overall power is increased relative to
the $\lambda_q=0$ case (the lowest curve); this is due to the decrease
in the denominator in eq.(\ref{eq:dimlesspowerspec}). We also see that
the $\lambda_q=0$ line acts as a lower envelope for all the
curves. This should not be a surprise since for those times for which
$\cos^2\left(\theta_{k f}(\eta)+\theta_{k g}(\eta)\right)=0$ we just
get the result of the uncoupled case back. At higher $k$ the
oscillation frequency becomes greater and greater such that
observations will not be able to resolve these oscillations. At lower
$k$ however, the oscillations have not yet coalesced and might give
rise to observable effects.

\begin{figure}[!htbp]
\begin{center}
\includegraphics[scale=0.6]{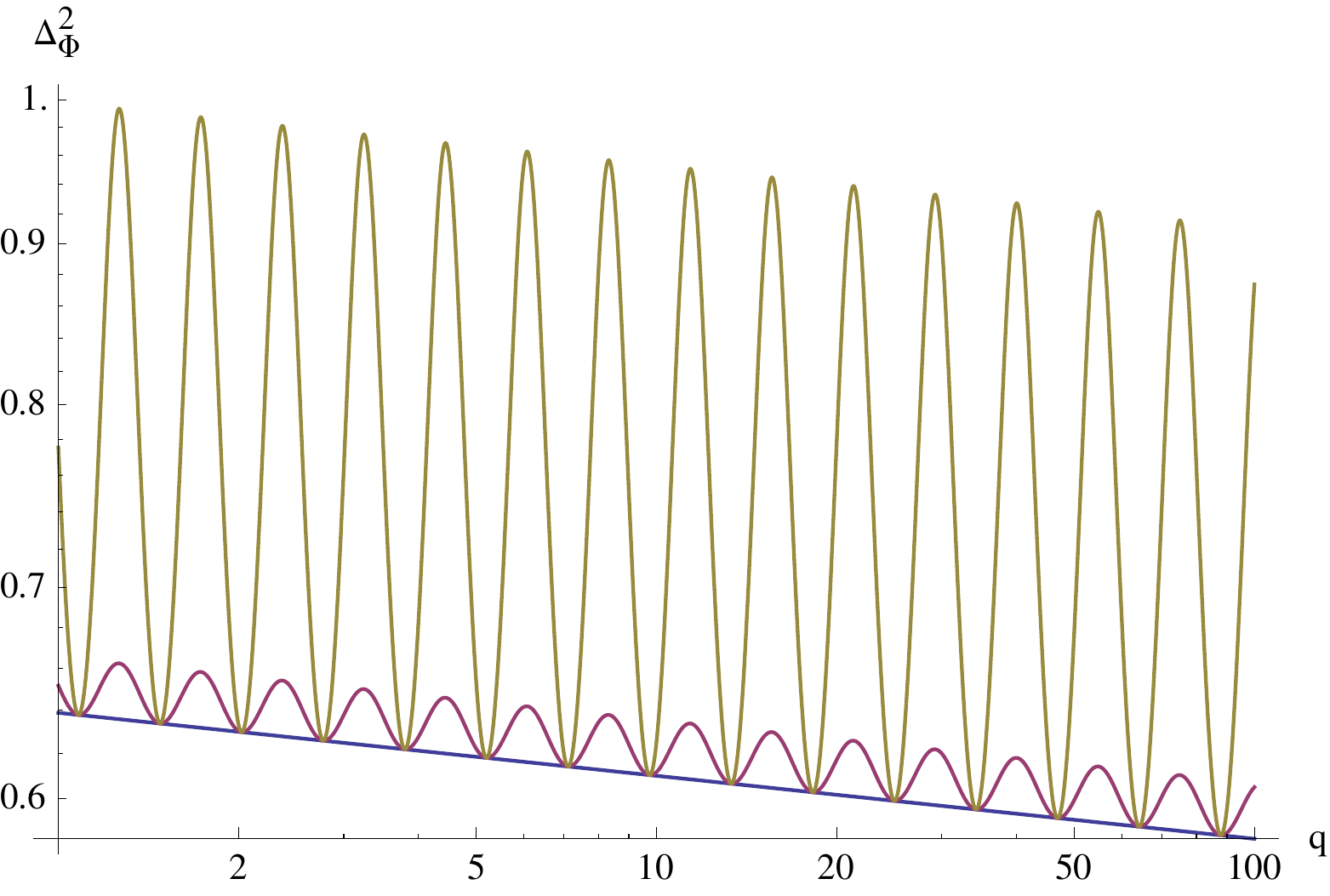}
\caption{A power spectrum plot with the same parameters as
  Fig.~\ref{fig:epsilon} except with a significant mass
  assigned to $\chi$: $m_{\chi}= 10 H$. The overall effects are similar but
  with different detailed features in the oscillations.}
\label{fig:massive}
\end{center}
\end{figure}

We can also consider the case where $\chi$ is massive with a mass
larger than $H$. In fig.(\ref{fig:massive}), we take $\nu_{\Phi} =
3\slash 2+\epsilon$ so that $\Phi$ is nearly massless and
$m_{\chi}\sim 10 H$. We see that in this case the phase of the
oscillations remains constant with $k$.  

\subsubsection{\label{subsubsec:perturbative} Perturbative Power Spectrum}

How accurate is our perturbative approximation to the modes and the power spectrum? If we consider the case where both fields are massless and minimally coupled, then using eq.(\ref{eq:field2ptfch}) we can infer the fractional change in the power spectrum $P(k) =  \langle \phi_{\vec{k}} \phi_{-\vec{k}} \rangle(\eta)$ relative to its unperturbed BD value:
\begin{eqnarray}
\label{eq:rel2pt}
& & \frac{\Delta P(k)}{P_{BD}(k)} = \lambda_k^2\left(2\ {\rm Re}{\cal F}_k(\eta)+ 4\cos^2\left(\theta^{BD}_{k f}+\theta^{BD}_{k g}\right)\right) = \\
& = & \lambda_k^2 \left(\cos 2 G(k\eta)+\cos 2 G(k \eta_0)-2\cos(G(k\eta)+G(k\eta_0)) + 4\cos^2 2(k\eta+\tan^{-1}\frac{1}{k\eta})\right).\nonumber
\end{eqnarray}
If we evaluate this at late times $\eta\rightarrow 0^-$ we arrive at
\begin{equation}
\label{eq:latetimerel2pt}		
\left . \frac{\Delta P(k)}{P_{BD}(k)}\right |_{\eta\rightarrow 0^-}=\lambda_k^2\left(4-2\sin^2 G(k\eta_0)+2 \sin G(k\eta_0)\right).
\end{equation}
In terms of our dimensionless variables defined above we have:
\begin{equation}
\label{eq:perturbdimeless}
\tilde{\Delta}^2_{\Phi\ \rm {pert}}(q) = \frac{1}{2}\left(1+\lambda_q^2\left(3+ \cos 2 G(q) + 2 \sin G(q)\right)\right).
\end{equation}

\begin{figure}[!htbp]
\begin{center}
\includegraphics[scale=0.6]{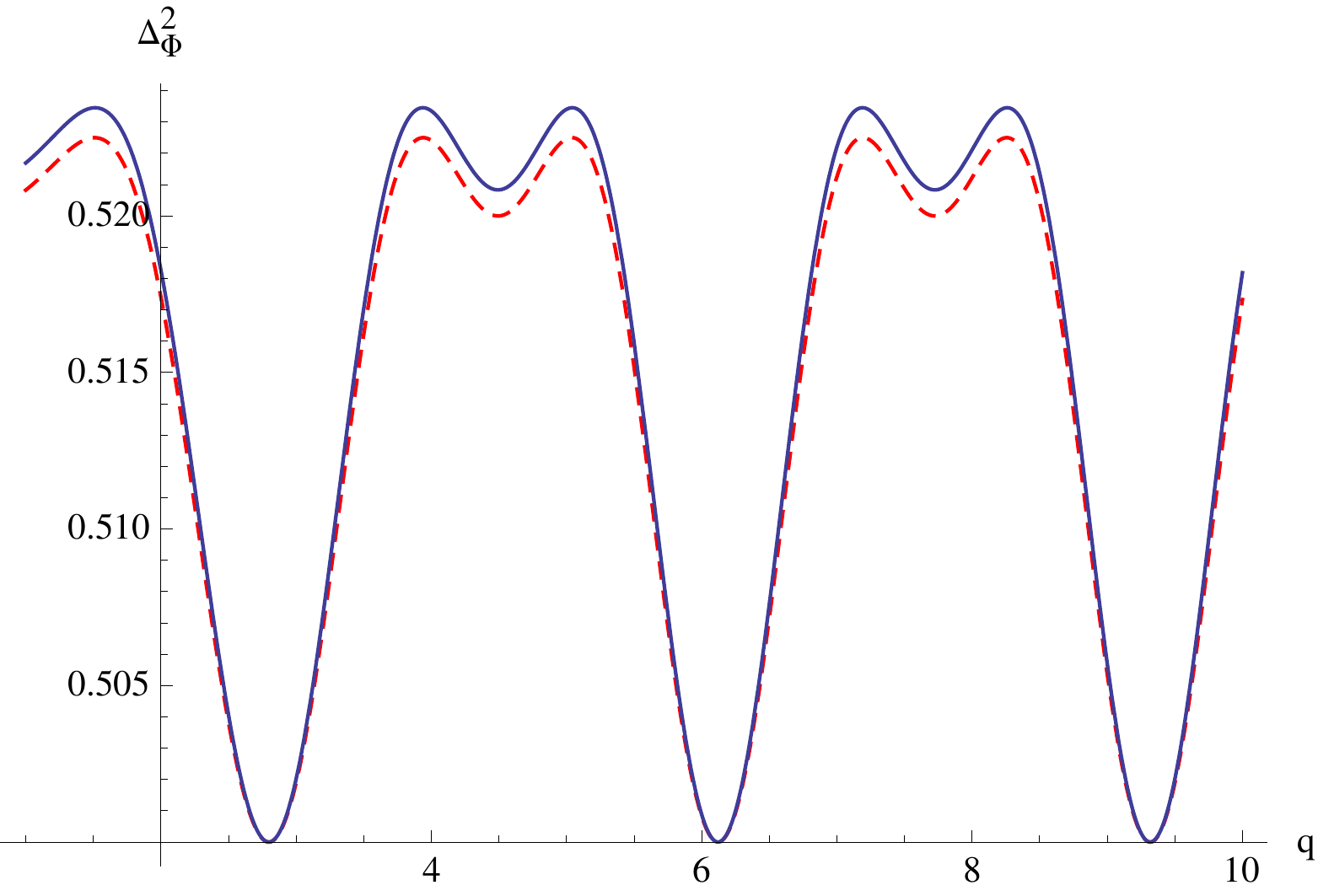}
\caption{Plots of $\tilde{\Delta}^2_{\Phi}(q)$ versus $q$, for $\lambda_q = 0.1$ and  $\nu_{\Phi}=\nu_{\chi}= 3\slash 2$, comparing the numerical solution (solid line) to the perturbative solution (dashed line).}
\label{fig:perturb}
\end{center}
\end{figure}

We plot both the perturbative as well as the numerical power spectra for this case with $\lambda_q=0.1,\ \nu_{\Phi}=\nu_{\chi}=3\slash 2$ in fig.(\ref{fig:perturb}). We see that for this case, the perturbative result captures both the qualitative behavior as well as most of the quantitative results compared to the numerical solution. 

More generally, we can do the integrals in eq.(\ref{eq:finalperts}) numerically to construct the perturbative modes and use those to compute the power spectrum given in eq.(\ref{eq:field2ptfch}).  

\begin{figure}[!htbp]
\begin{center}
\includegraphics[scale=0.6]{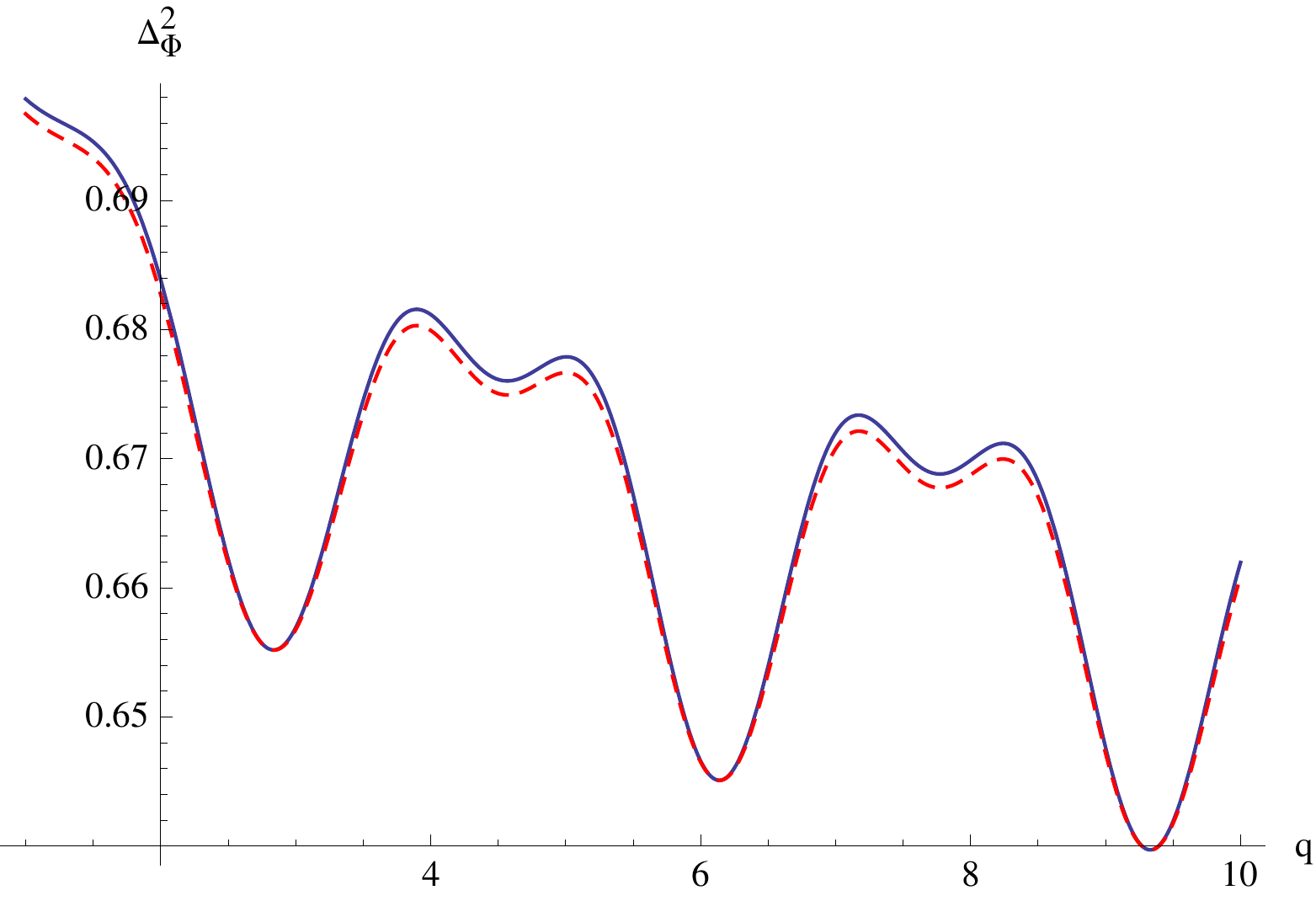}
\caption{Plots of $\tilde{\Delta}^2_{\Phi}(q)$ versus $q$, for $\lambda_q = 0.1$ $\nu_{\Phi}=3\slash 2 +0.01,\ \nu_{\chi}= 3\slash 2$, comparing the numerical solution (solid line) to the perturbative solution (dashed line).}
\label{fig:perturbcomparisonfull}
\end{center}
\end{figure}

\begin{figure}[!htbp]
\begin{center}
\includegraphics[scale=0.6]{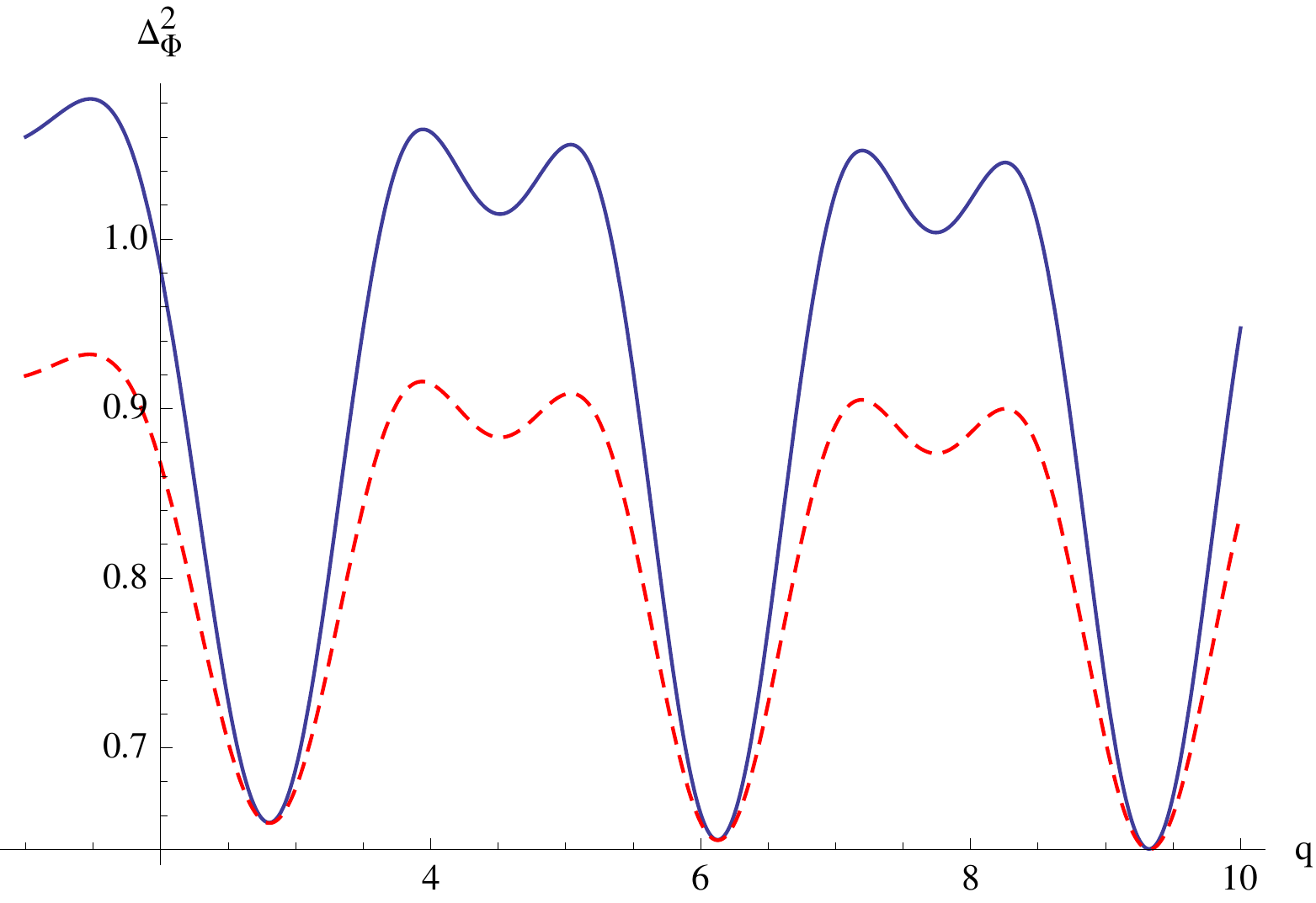}
\caption{Plots of $\tilde{\Delta}^2_{\Phi}(q)$ versus $q$, for $\lambda_q = 0.3$ $\nu_{\Phi}=3\slash 2 +0.01,\ \nu_{\chi}= 3\slash 2$, comparing the numerical solution (solid line) to the perturbative solution (dashed line). The agreement is significantly worse than at $\lambda_q=0.1$.}
\label{fig:comparison00103}
\end{center}
\end{figure}

Fig.(\ref{fig:perturbcomparisonfull}) shows that for  $\lambda_q=0.1$ there is very good agreement between the approximate and full numerical solutions. Needless to say, this agreement gets worse as $\lambda_q$ increases. 
For $\lambda_q=0.3$, we plot both curves in fig.(\ref{fig:comparison00103}). We see that by this point the difference between the two curves is large enough that we should not trust the perturbative solution here. 

\section{\label{sec: conclusions} Conclusions and Further Directions}

Our goal in this work was to exploit the fact that the inflaton would
most likely not appear alone, but accompanied by a plethora of other
scalar fields, all of which would also be in some quantum state at an
initial time $\eta_0$. Thinking of entangled states of these fields
with the inflaton is then quite natural. What we've seen here is that
even in the case where the fields are completely decoupled from each
other, initial state entanglement couples the modes together through
the Schr\"odinger equation for the wave functional describing the
system. This coupling of the modes then modifies cosmological
observables such as the power spectrum and presumably would also have
an effect on the bispectrum as well. In the example we worked out
here, with two free coupled scalar fields, we saw that the standard
constraints on modifications of the quantum state of inflaton
fluctuations, such as the back reaction constraint, could be satisfied
while still allowing for potentially observable deviations from the BD
results for the power spectrum.  

Perhaps the simplest way to describe
what we have done is that we have used the prospect of entanglement
with additional fields to motivate choosing a state for the
inflaton other than the usual Bunch Davies (BD) vacuum. The entanglement we
consider puts the inflaton in a mixture of different energy eigenstates, not
just the single BD state.  Having constructed such a state it is not
surprising that oscillations appear, which is a generic outcome when
states are made of simple combinations of different energy
eigenstates.

%% Entangling the fields together introduced oscillations into their
%% power spectra. These oscillations are akin to what happens with
%% massive neutrinos. Introducing an entangled initial state into the
%% evolution of the density matrix as in ref.\cite{Agarwal:2012mq} is the
%% same as inserting mixing into the mass matrix of the fields. This then
%% gives rise to oscillations in time between the various two-point
%% functions of the fields in theory. 

The size of the oscillations induced by the entanglement depends
strongly on the value of $\lambda_k$; this should give us a large
lever arm that we can use constrain large portions of the entanglement
parameter space using currently available data. What is needed next is
to generalize our calculations to the case where the near massless
field is the curvature fluctuation $\zeta(\vec{x},t)$ entangle it with
some other scalar field and then run the resulting power spectra
through a Boltzmann integrator that can deal with oscillatory
primordial power spectra\cite{Meerburg:2013cla,Meerburg:2013dla} to
construct the resulting $C_l$'s; this will appear in later
work. In particular, if the BICEP2 results are determined to be
cosmological~\cite{Ade:2014xna} there are tensions between it and the Planck
data on the tensor to scalar amplitude ratio $r$. So it might be
interesting to entangle $\zeta$ with the (also gauge invariant) tensor
perturbations $h_{i j}$. This would modify both contributions to the
TT power spectrum and introduce a dependence on the entanglement
parameter $\lambda_k$ into the Planck bounds on $r$.

We can also use our density matrix together with the cubic
interactions of $\zeta$ to compute the effects of this new state on
the
bispectrum\cite{Holman:2007na,Ganc:2011dy,Chen:2006nt,Agullo:2010ws,Agarwal:2012mq}
and in particular, check the so-called consistency
relation\cite{Maldacena:2002vr,Creminelli:2004yq,Cheung:2007sv}.  

There are other calculations to be done but now dealing with the
evolution of the inflaton zero mode. We can generalize our entangled
state to allow for a $\Phi$ zero mode and then use the tadpole
condition\cite{Boyanovsky:1994me} to understand how the self
consistent equation of motion for the zero mode (including the
semiclassical back reaction of the fluctuations on the FRW geometry)
gets modified by the entanglement. Also, it would be interesting to
revisit the issues involved in reheating now that a new type of
coupling between fields is allowed. 
 
The question of exactly {\em how} an entangled initial state might be produced is also a fascinating one and should be investigated further (for some work on entangled states in de Sitter space, see \cite{Lello:2013qza}).

In short, quantum states where the inflaton fluctuations are entangled
with those of other fields offer a new set of vistas to be
explored. We find it very interesting that current data may already place
interesting bounds on how much entanglement could have occurred in the
initial state, and we expect that many new effects will appear as our
understanding of these states deepens. 

\acknowledgments  We thank N.~Kaloper and L.~Knox for useful
communications. R.~H. was supported in part by the Department of 
Energy under grant DE-FG03-91-ER40682, as well as by a grant from the
John Templeton Foundation. He would also like to thank the Physics
Department at UC Davis for hospitality while this work was in
progress. A.~A. and N.~B. were supported in part by DOE Grants
DE-FG02-91ER40674 and DE-FG03- 
91ER40674 and the National Science Foundation under Grant No. PHY11-25915.


\bibliographystyle{JHEP}
\bibliography{cosmoentanglement}
\end{document}